\documentclass[runningheads]{llncs}

\usepackage{eccv}

\usepackage{algorithm}
\usepackage{algpseudocode}

\usepackage{tikz}
\usetikzlibrary{calc,trees,positioning,arrows,chains,shapes.geometric,%
    decorations.pathreplacing,decorations.pathmorphing,shapes,%
    matrix,shapes.symbols}

\usepackage{pgfplots}
\usepgfplotslibrary{groupplots}
\pgfplotsset{
    yticklabel style={
        /pgf/number format/fixed,
        /pgf/number format/precision=8
    },
    scaled y ticks=false,
    xticklabel style={
        /pgf/number format/fixed,
        /pgf/number format/precision=8
    },
    xlabel style={
        yshift=0.1cm,
        font=\scriptsize
    },
    scaled x ticks=false,
    legend style={
        /tikz/every even column/.append style={column sep=0.2cm}
    },
    height=4.5cm,
}

\usepackage{eccvabbrv}

\usepackage{graphicx}
\usepackage{booktabs}

\usepackage[accsupp]{axessibility}  

\usepackage{hyperref}

\usepackage{orcidlink}

\tikzset{
roundrect/.style={
    rectangle, 
    rounded corners, 
    draw=black, very thick,
    text width=8em, 
    minimum height=3em, 
    text centered},
sroundrect/.style={
    rectangle, 
    rounded corners, 
    draw=black, very thick,
    text width=6em, 
    minimum height=3em, 
    text centered},
rect/.style={
    rectangle, 
    draw=black, very thick,
    text width=5em, 
    minimum height=5em, 
    text centered},
img/.style={
    inner sep=0,
    outer sep=0,
    text width=5em, 
    minimum height=5em, 
    text centered},
every join/.style={->},
}

\newcommand{\comment}[1]{}

\begin{document}

\title{Higher fidelity perceptual image and video compression with a latent conditioned residual denoising diffusion model} 
\titlerunning{Image compression with a latent conditioned residual diffusion model}

\author{Jonas Brenig\orcidlink{0009-0004-3770-9643} \and
Radu Timofte\orcidlink{0000-0002-1478-0402}}

\authorrunning{J.~Brenig, R.~Timofte}
\institute{Computer Vision Lab, CAIDAS \& IFI, University of W\"urzburg, Germany}

\maketitle

\begin{abstract}
    Denoising diffusion models achieved impressive results on several image generation tasks often outperforming GAN based models. 
    Recently, the generative capabilities of diffusion models have been employed for perceptual image compression, such as in CDC~\cite{yang2024lossy}. A major drawback of these diffusion-based methods is that, while producing impressive perceptual quality images they are dropping in fidelity/increasing the distortion to the original uncompressed images when compared with other traditional or learned image compression schemes aiming for fidelity. 
    
    In this paper, we propose a hybrid compression scheme optimized for perceptual quality, extending the approach of the CDC model with a decoder network in order to reduce the impact on distortion metrics such as PSNR. After using the decoder network to generate an initial image, optimized for distortion, the latent conditioned diffusion model refines the reconstruction for perceptual quality by predicting the residual. On standard benchmarks, we achieve up to +2dB PSNR fidelity improvements while maintaining comparable LPIPS and FID perceptual scores when compared with CDC. Additionally, the approach is easily extensible to video compression, where we achieve similar results.
\end{abstract}

\section{Introduction}
\label{sec:intro}

With ever-increasing resolutions of photo cameras, good compression schemes remain an important to reduce bandwidth and storage requirements for images.
While there are some newer hand-crafted codecs, such as BPG~\cite{bpg},  which improve drastically over older codecs such as JPEG~\cite{wallace1991jpeg}, learned approaches have been shown to achieve significantly lower bitrates at higher quality. 

Most neural network based image compression methods use an autoencoder with some kind of learned entropy model \cite{balleVariationalImageCompression2018,minnenJointAutoregressiveHierarchical2018,he2022elic}. 
Recently, there has been some interest in optimizing compression models for perceptual quality instead of optimizing solely for distortion metrics. Most of these models improve perceptual quality by incorporating a perceptual distance metrics, such as LPIPS~\cite{zhang2018unreasonable}, as an additional loss during training. Notably, in combination with adversarial neural networks (GAN), this approach was shown to significantly improve perceptual quality \cite{mentzerHighFidelityGenerativeImage2020, he2022po}.

Diffusion models and related techniques are known to be able to generate high quality samples \cite{hoDenoisingDiffusionProbabilistic2020, song2020score, rombach2021high}, surpassing even the quality of generative adversarial networks \cite{dhariwalDiffusionModelsBeat2021} for a number of generative tasks. Additionally, they are generally easier to train as they do not suffer from the same stability problems often encountered when working with GANs.
Despite their success for other generative tasks, diffusion models are still rarely used in the context of image compression.

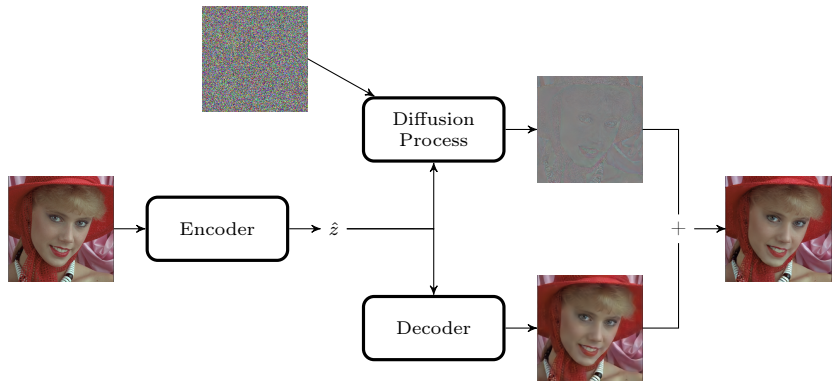
\begin{figure}[!th]
    \centering
    \scriptsize
    \begin{tikzpicture}[node distance=1.5em, >=stealth']
        \node[img] (input) {\includegraphics[width=5em]{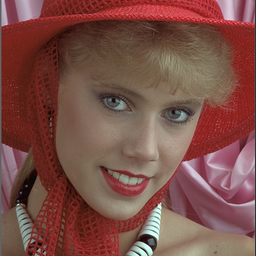}};
        \node[sroundrect, right=of input] (encoder) {Encoder};
        \node[right=of encoder] (latent) {$\hat z$};
        \node[right=0em of latent] (r1) {};
        \node[below=4em of r1.south] (r2) {};
        \node[above=4em of r1.north] (r3) {};
        \node[above left=3em and 4em of r1] (r4) {};
        \node[img, above = of r4] (noise) {\includegraphics[width=5em]{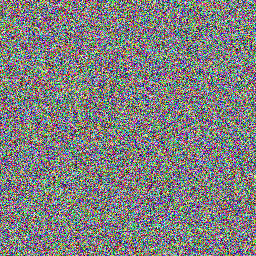}};
        \node[sroundrect, right=0em of r2] (decoder) {Decoder};
        \node[sroundrect, right=0em of r3] (diffusion) {Diffusion Process};
        \node[img, right=of decoder] (dec_recon) {\includegraphics[width=5em]{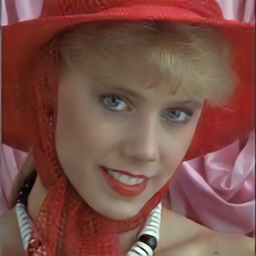}};
        \node[img, right=of diffusion] (residual) {\includegraphics[width=5em]{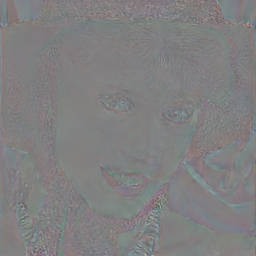}};
        \node[right=15em of latent] (t1) {+};
        \node[img, right=of t1] (final_image) {\includegraphics[width=5em]{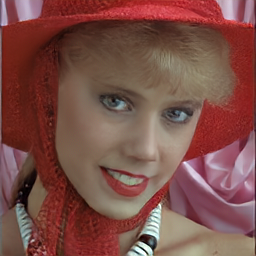}};
        
        \draw [ ->] (input.east) -- (encoder.west);
        \draw [ ->] (encoder.east) -- (latent);
        \draw [ ->] (noise.east) -- (diffusion);
        \draw [ ->] (latent.east) -| (diffusion);
        \draw [ ->] (latent.east) -| (decoder);
        \draw [ ->] (decoder.east) -- (dec_recon);
        \draw [ ->] (diffusion.east) -- (residual);
        \draw [ ] (dec_recon.east) -| (t1.south);
        \draw [ ] (residual.east) -| (t1.north);
        \draw [ ->] (t1.east) -- (final_image);        
    \end{tikzpicture}
    \vspace{-0.5em}
    \caption{Overview of the proposed architecture}
    \label{fig:architecture}
\end{figure}
There has been a number of works in closely related generative tasks such as superresolution or image restoration~\cite{saharia2022image,kawarDenoisingDiffusionRestoration2022,kawarJPEGArtifactCorrection2022,chung2022improving}. 
Diffusion models have also been used to improve the perceptual quality of pretrained neural image compression methods \cite{ghouseResidualDiffusionModel2023, hoogeboom2023high}.
However, existing end-to-end trained diffusion models for compression are either only applicable to smaller images \cite{theisLossyCompressionGaussian2022} or score low on fidelity metrics \cite{yang2024lossy}. 

\begin{figure*}[!htb]
    \centering
    \renewcommand{\arraystretch}{0}
    \setlength{\tabcolsep}{0pt}
    
    \begin{tabular}{ccc}
         \subfloat[][Ground Truth]{\includegraphics[width=.33\textwidth]{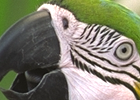}} &
         \subfloat[][CDC $\epsilon$ ($\rho\!=\!0.9$) \tiny 0.160bpp]{\includegraphics[width=.33\textwidth]{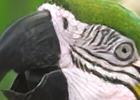}} &
         \subfloat[CDC $x_0$ ($\rho\!=\!0.9$) \tiny 0.205bpp]{\includegraphics[width=.33\textwidth]{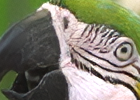}} \\
         \subfloat[\textbf{ours} Decoder \tiny 0.157bpp]{\includegraphics[width=.33\textwidth]{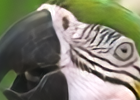}} &
         \subfloat[\textbf{ours} ($\rho\!=\!0.1$) \tiny 0.136bpp]{\includegraphics[width=.33\textwidth]{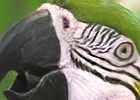}} &
         \subfloat[\textbf{ours} ($\rho\!=\!0.5$) \tiny 0.157bpp]{\includegraphics[width=.33\textwidth]{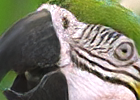}} \\[-2ex]
    \end{tabular}
    \vspace{1ex}
    \caption{Qualitative comparison between CDC and ResCDC (ours)}\label{fig:qualitative}
\end{figure*}

In this work, we propose a hybrid approach based on an autoencoder-based compression method which uses a conditional diffusion model to optimize the reconstruction for perceptual quality.
Extending on the ideas of CDC~\cite{yang2024lossy}, we introduce an end-to-end trained hybrid approach that utilizes a traditional decoder as well as a diffusion model which is conditioned using the same latent as the decoder.
While the decoder is trained using the $l_2$-loss the diffusion model is optimized for perceptual quality.

The diffusion model is trained to predict the residuals, following Whang \etal who found this approach to speed up sampling of their deblurring model~\cite{whang2022deblurring}. Using this approach we are able to improve the fidelity of the diffusion based reconstruction using only a few sampling steps.

The approach can easily be adjusted to a video compression or enhancement setting as well. 
In Section~\ref{sec:video} we show how a simple adaptation can improve perceptual quality of existing neural video compression codecs by training the diffusion model on top of a pretrained autoencoder based model.

\section{Related Work}

\subsection{(Lossy) Image Compression}
Lossy image compression is usually done using hand-crafted algorithms such as the widely used JPEG~\cite{wallace1991jpeg}. More modern approaches like BPG~\cite{bpg} are able to achieve much better results. However, during recent years neural compression techniques were able to surpass hand-crafted algorithms regarding perceptual quality, as well as distortion metrics such as PSNR.

Most approaches are based on an encoder / decoder setup in which the encoder latent is quantized, and then compressed using a lossless arithmetic coder. 
The very popular approach by Ballé \etal \cite{balleVariationalImageCompression2018} is based on a simple autoencoder, with an additional hyper-prior model.
The hyper-prior is used to learn a good probability model to allow better compression of the latent. 
Newer approaches generally try to extend this idea, often extending or augmenting the hyper-prior in some way \cite{minnenJointAutoregressiveHierarchical2018, he2022elic}.

When optimizing for perceptual quality GAN-based approaches have been very successful. 
For example, HiFiC~\cite{mentzerHighFidelityGenerativeImage2020} extends the autoencoder-based approach, by a GAN training procedure and an additional perceptual loss component. Some newer approaches usually combine new ideas from autoencoders with adversarial losses to improve upon HiFiC \cite{he2022po}.

\subsection{Diffusion Models for Image Enhancement}

Diffusion models have been successfully applied to a variety of image enhancement tasks, including superresolution~\cite{saharia2022image, rombachHighResolutionImageSynthesis2022} and compressed image enhancement~\cite{ghouseResidualDiffusionModel2023,hoogeboom2023high}.
Ghouse \etal \cite{ghouseResidualDiffusionModel2023} propose to use a conditional diffusion model to enhance a pretrained base-codec. The diffusion model uses the initial reconstruction in a conditional diffusion model which predicts the residual to enhance the final output. While conceptually somewhat similar to our proposed method, the method is not trained end-to-end. Therefore, it might be better classified as a method for (neural) compressed image enhancement.
Another very similar approach by Hoogeboom \etal\cite{hoogeboom2023high} also used a conditional diffusion conditioned on an initial reconstruction of an already strong pretrained codec, to enhance the final image. However, instead of predicting a residual, they predicted patches of the final image directly.

Similarly, there are works using diffusion models for JPEG artifact correction such as \cite{kawarJPEGArtifactCorrection2022, saharia2022palette}. 

\subsection{Diffusion Models for Image Compression}
There have been a number of works which try to employ diffusion models directly for image compression.

Theis \etal \cite{theisLossyCompressionGaussian2022} propose to use unconditional diffusion models in combination with reverse channel coding to provide a compression scheme based on the communication of noisy samples. However, this approach is quite expensive computationally due to the reliance on the very slow reverse channel coding process. Therefore, they were only able to experiment on images with a resolution of 64x64.

In this paper we build upon CDC of Yang and Mandt \cite{yang2024lossy} who use conditional diffusion models to replace the traditional decoder. 
While the diffusion process incurs additional cost during decoding, they were able to achieve very competitive results when evaluating on a variety of perceptual metrics. 

Other approaches have shown how diffusion models might be used for lossless compression \cite{hoogeboom2021autoregressive}, or for extreme low bitrate scenarios via text embeddings \cite{panExtremeGenerativeImage2022}.

\subsection{Video Compression and Enhancement}

Compressing video brings its own challenges concerning temporal consistency and encoding / decoding speed. While most learned video compression methods fundamentally work similar to learned image compression methods, they require additional components to capture the temporal nature of video \cite{lu2019dvc, hu2021fvc, agustsson2020scale, li2021deep,li2024neural}. 
Some methods also use recurrent network components to improve the use of temporal information \cite{yang2020learning,yang2022advancing,li2024neural}.

In this paper, we experimented with applying our approach on top of learned video compression methods such as Scale-Space Flow~\cite{agustsson2020scale}, DVC~\cite{lu2019dvc} and DCVC-FM~\cite{li2024neural} by incorporating their latent as conditioning information for the diffusion unet.

Existing learned perceptual video compression methods usually use the same combination of perceptual losses and GANs as is the case for image compression models \cite{yang2022perceptual, mentzer2022neural}.

There has been some works that try to enhance compressed video (usually focused on traditional codecs such as H265) to achieve better perceptual quality \cite{zheng2022progressive, chan2022basicvsr++}.

Diffusion models do not yet see much use in this area, probably mostly due to their slow sampling speed. Nonetheless, they have been used in the context of video enhancement and/or superresolution \cite{zhou2024upscale, zhou2022cadm} 
They have also been used for extreme video compression \cite{li2024extreme}, where pretrained diffusion models were used to predict new frames, skipping encoding of some frames.

\section{Method}
Our method extends the CDC~\cite{yang2024lossy} by using an additional decoder network.
Similarly, it builds on top of an auto-encoder based approach like the MS-Hyperprior \cite{minnenJointAutoregressiveHierarchical2018} and uses a denoising unet similar to DDIM and DDPM \cite{songDenoisingDiffusionImplicit2022, hoDenoisingDiffusionProbabilistic2020}.
However, for decoding we use a decoder network in combination with a residual denoising diffusion model conditioned on the encoder latent (as illustrated in Figure \ref{fig:architecture}).

Similar to \cite{whang2022deblurring} and \cite{ghouseResidualDiffusionModel2023}, the diffusion model tries to predict the difference $r$ between the decoder reconstruction and the original image.
The key motivation here is that residuals are simpler to model since the pixels roughly follow a normal distribution, which can reduce the number of sampling steps required~\cite{whang2022deblurring,ghouseResidualDiffusionModel2023}.
Note, that the decoder is trained jointly with the diffusion model, which means that both components of the decoding model influence latent generated by the encoder during training. This separates our approach from some other diffusion based methods for enhancing perceptual quality of an autoencoder based codec~\cite{ghouseResidualDiffusionModel2023,hoogeboom2023high}.

In practice, the decoder will produce a fidelity-optimized reconstruction, while the latent conditioned residual diffusion model adds additional detail, finetuning the image for perceptual quality.
\begin{algorithm}[hbt!]
    \caption{Training}\label{alg:training}
    \begin{algorithmic}
        \Require $N$ \Comment{Number of training diffusion steps}
        \Loop
            \State $x \gets \text{batch}$, $t \sim \mathcal{U}(0, N)$, $\epsilon \sim \mathcal{N}(0, 1)$

            \State $\hat{y}, \hat{z} = enc(x)$ \Comment{Pseudo-quantized encoder output (latent and hyper-latent)}
            \State $\hat{x} = dec(\hat{y})$ \Comment{Decoder reconstruction}

            \State $r = x - \hat{x}$ \Comment{Calculate residual}
            
            \State $r_t = \sqrt{\alpha_t} r + \sqrt{1-\alpha_t}\epsilon$ \Comment{Diffusion forward step}
            \State $\hat{\epsilon} = f(r_t, \hat{y}, t / N)$ \Comment{Noise predicting model}

            \State $\mathcal{L}_{dist} = \lvert\lvert\epsilon - \hat{\epsilon}\rvert\rvert_1$
            \State $\mathcal{L}_{perc} = LPIPS(x, \hat{x} + \hat{r}_0)$
            \State $\mathcal{L}_{decoder} =  \lvert\lvert x - \hat{x}\rvert\rvert^2_2$
            
            \State $\mathcal{L}_{bitrate} = \log P(\hat{y}) + \log P(\hat{z})$
            
            \State $\mathcal{L} = (1-\rho)\mathcal{L}_{dist}  + \rho \mathcal{L}_{perc} + \lambda \mathcal{L}_{bitrate} + \mathcal{L}_{decoder}$

            \State \textrightarrow \textit{optimizer step}
            
        \EndLoop
    \end{algorithmic}
\end{algorithm}
\begin{algorithm}[hbt!]
    \caption{Decompression}\label{alg:sampling}
    \begin{algorithmic}
        \Require $N$ \Comment{Number of sampling steps}
        \Require $\hat{y}$ \Comment{Compressed latent}
        
        \State $t \gets N$
        \State $r_t = 0$
        \While{$t > 0$}
            \State $t \gets t - 1$
            \State $\hat{\epsilon}_t = f(r_t, \hat{y}, t / N)$ \Comment{Noise predicting model}

            \State $r_t = DDIM(\hat{\epsilon}_t, x_r, t)$ \Comment{Diffusion reverse step}
            
        \EndWhile

        \State $\hat{x} = dec(\hat{y})$ \Comment{Decoder reconstruction}
        \State $x_{rec} = \hat{x} + r_t$ \Comment{Combine with residual for final image}
    \end{algorithmic}
\end{algorithm}
\pgfplotscreateplotcyclelist{main}{
    black\\%
    red,mark=x\\%
    blue,mark=+\\%
    teal,mark=*\\%
    orange,mark=triangle*\\%
    pink,mark=o\\%
}
\begin{figure*}[!htb]
    \centering
    \begin{tikzpicture}
        \begin{groupplot}[
                legend style={at={(0.5,1.01)}, anchor=south, nodes={scale=0.8, transform shape}},
                grid=both,
                grid style={line width=.1pt, draw=gray!10},
                group style={
                    group size=3 by 2,
                    vertical sep=3em
                },
                legend columns=3,
                cycle list name=main,
                legend to name=leg-res-main,
                title style={yshift=-0.5em}, 
                height=4.5cm,
                width=.38\textwidth,
                xmin=0.12,
                xmax=0.8,
                xlabel={bits per pixel (bpp)},
            ]
            \nextgroupplot[
                title=PSNR \textuparrow,
                ytick distance=2,
            ]
            \addplot table [x=bpp, y=psnr, col sep=comma] {csv/bpg_0.csv};
            \addplot table [x=BPP, y=PSNR, col sep=comma] {csv/cdc_epsilon_rho09.csv};
            \addplot table [x=BPP, y=PSNR, col sep=comma] {csv/cdc_x0_rho09.csv};
            \addplot table [x=BPP, y=PSNR, col sep=comma] {csv/cdc_residual_epsilon_rho05.csv};
            \addplot table [x=BPP, y=PSNR, col sep=comma] {csv/cdc_residual_epsilon_rho01.csv};
            \addplot table [x=bpp, y=psnr, col sep=comma] {csv/HiFiC-PyTorch_kodak.csv};

            \nextgroupplot[
                title=LPIPS \textdownarrow,
                ymax=0.13,
            ]            
            \addplot table [x=bpp, y=lpips, col sep=comma] {csv/bpg_0.csv};
            \addplot table [x=BPP, y=LPIPS, col sep=comma] {csv/cdc_epsilon_rho09.csv};
            \addplot table [x=BPP, y=LPIPS, col sep=comma] {csv/cdc_x0_rho09.csv};
            \addplot table [x=BPP, y=LPIPS, col sep=comma] {csv/cdc_residual_epsilon_rho05.csv};
            \addplot table [x=BPP, y=LPIPS, col sep=comma] {csv/cdc_residual_epsilon_rho01.csv};
            \addplot table [x=bpp, y=lpips, col sep=comma] {csv/HiFiC-PyTorch_kodak.csv};
            
            \nextgroupplot[
                title=MS-SSIM \textuparrow,
                ytick distance=0.02,
            ]                 
            \addplot table [x=bpp, y=ms-ssim, col sep=comma] {csv/bpg_0.csv};
            \addplot table [x=BPP, y=MS-SSIM, col sep=comma] {csv/cdc_epsilon_rho09.csv};
            \addplot table [x=BPP, y=MS-SSIM, col sep=comma] {csv/cdc_x0_rho09.csv};
            \addplot table [x=BPP, y=MS-SSIM, col sep=comma] {csv/cdc_residual_epsilon_rho05.csv};
            \addplot table [x=BPP, y=MS-SSIM, col sep=comma] {csv/cdc_residual_epsilon_rho01.csv};
            \addplot table [x=bpp, y=ms-ssim, col sep=comma] {csv/HiFiC-PyTorch_kodak.csv};
            
            \legend {BPG, CDC $\epsilon$ ($\rho=0.9$), CDC $x_0$ ($\rho=0.9$), ResCDC ($\rho=0.5$),  ResCDC ($\rho=0.1$), HiFiC}
        \end{groupplot}
    \end{tikzpicture}

    \vspace{0cm}
    \pgfplotslegendfromname{leg-res-main}
    \caption{Results on the Kodak images. CDC $\epsilon$ models were sampled using 500 steps, CDC $x_0$ models using 17 steps, ResCDC (ours) models using 100 steps.}
    \label{fig:eval:kodak}
\end{figure*}
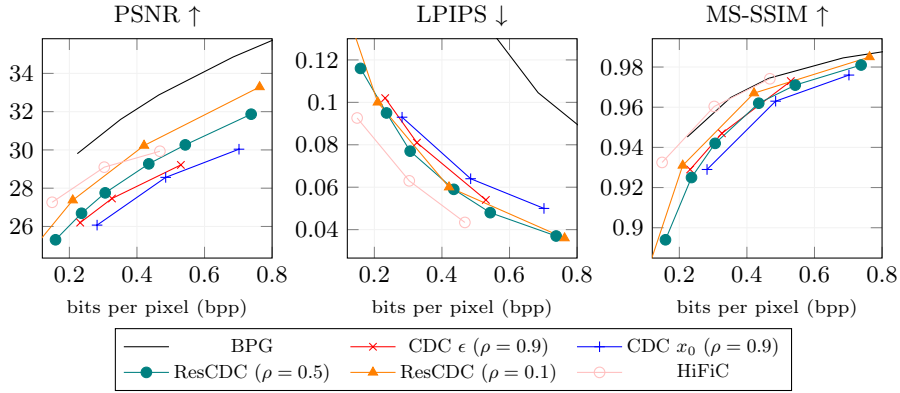

\begin{figure*}[!htb]
    \centering
    \begin{tikzpicture}
         \begin{groupplot}[
                legend style={at={(0.5,1.01)}, anchor=south, nodes={scale=0.8, transform shape}},
                grid=both,
                grid style={line width=.1pt, draw=gray!10},
                group style={
                    group size=2 by 2,
                    vertical sep=3em
                },
                legend columns=3,
                cycle list name=main,
                legend to name=leg-res-div2k,
                title style={yshift=-0.5em}, 
                width=.53\textwidth,
                xmax=0.8,
            ]
            \nextgroupplot[
                title=PSNR \textuparrow,
                ytick distance=2,
            ]           
            \addplot table [x=bpp, y=psnr, col sep=comma] {csv/div2k/bpg_0.csv};
            \addplot table [x=BPP, y=PSNR, col sep=comma, blue, mark=o] {csv/div2k/cdc_epsilon_rho09.csv};
            \addplot table [x=BPP, y=PSNR, col sep=comma] {csv/div2k/cdc_x0_rho09.csv};
            \addplot table [x=bpp, y=psnr, col sep=comma] {csv/div2k/cdc_residual_epsilon_rho05.csv};
            \addplot table [x=bpp, y=psnr, col sep=comma] {csv/div2k/cdc_residual_epsilon_rho01.csv};
            \addplot table [x=bpp, y=psnr, col sep=comma] {csv/div2k/HiFiC-PyTorch_div2k_768_pytorch.csv};

            \nextgroupplot[
                title=LPIPS \textdownarrow,
                xmax=0.8,
                ymax=0.12
            ]            
            \addplot table [x=bpp, y=lpips, col sep=comma] {csv/div2k/bpg_0.csv};           
            \addplot table [x=BPP, y=LPIPS, col sep=comma] {csv/div2k/cdc_epsilon_rho09.csv};
            \addplot table [x=BPP, y=LPIPS, col sep=comma] {csv/div2k/cdc_x0_rho09.csv};
            \addplot table [x=bpp, y=lpips, col sep=comma] {csv/div2k/cdc_residual_epsilon_rho05.csv};
            \addplot table [x=bpp, y=lpips, col sep=comma] {csv/div2k/cdc_residual_epsilon_rho01.csv};
            \addplot table [x=bpp, y=lpips, col sep=comma] {csv/div2k/HiFiC-PyTorch_div2k_768_pytorch.csv};
            
            \nextgroupplot[
                title=MS-SSIM \textuparrow,
            ]             
            \addplot table [x=bpp, y=ms-ssim, col sep=comma] {csv/div2k/bpg_0.csv};          
            \addplot table [x=BPP, y=MS-SSIM, col sep=comma] {csv/div2k/cdc_epsilon_rho09.csv};
            \addplot table [x=BPP, y=MS-SSIM, col sep=comma] {csv/div2k/cdc_x0_rho09.csv};
            \addplot table [x=bpp, y=ms-ssim, col sep=comma] {csv/div2k/cdc_residual_epsilon_rho05.csv};
            \addplot table [x=bpp, y=ms-ssim, col sep=comma] {csv/div2k/cdc_residual_epsilon_rho01.csv};
            \addplot table [x=bpp, y=ms-ssim, col sep=comma] {csv/div2k/HiFiC-PyTorch_div2k_768_pytorch.csv};

            \nextgroupplot[
                title=FID \textdownarrow,
                ymax=40,
                ytick distance=5,
            ]            
            \addplot table [x=bpp, y=fid, col sep=comma] {csv/div2k/bpg_0.csv};
            \addplot table [x=BPP, y=FID, col sep=comma] {csv/div2k/cdc_epsilon_rho09.csv};
            \addplot table [x=BPP, y=FID, col sep=comma] {csv/div2k/cdc_x0_rho09.csv};
            \addplot table [x=bpp, y=fid, col sep=comma] {csv/div2k/cdc_residual_epsilon_rho05.csv};
            \addplot table [x=bpp, y=fid, col sep=comma] {csv/div2k/cdc_residual_epsilon_rho01.csv};
            \addplot table [x=bpp, y=fid, col sep=comma] {csv/div2k/HiFiC-PyTorch_div2k_768_pytorch.csv};
            
            \legend {BPG, CDC $\epsilon$ ($\rho=0.9$), CDC $x_0$ ($\rho=0.9$), ResCDC ($\rho=0.5$),  ResCDC ($\rho=0.1$), HiFiC}
        \end{groupplot}
    \end{tikzpicture}
    \vspace{0cm}
    \pgfplotslegendfromname{leg-res-div2k}
    \caption{Results on the DIV2K validation-set. CDC $\epsilon$ models were sampled using 500 steps, CDC $x_0$ models using 17 steps, ResCDC (ours) models using 100 steps.}\label{fig:eval:div2k}
\end{figure*}
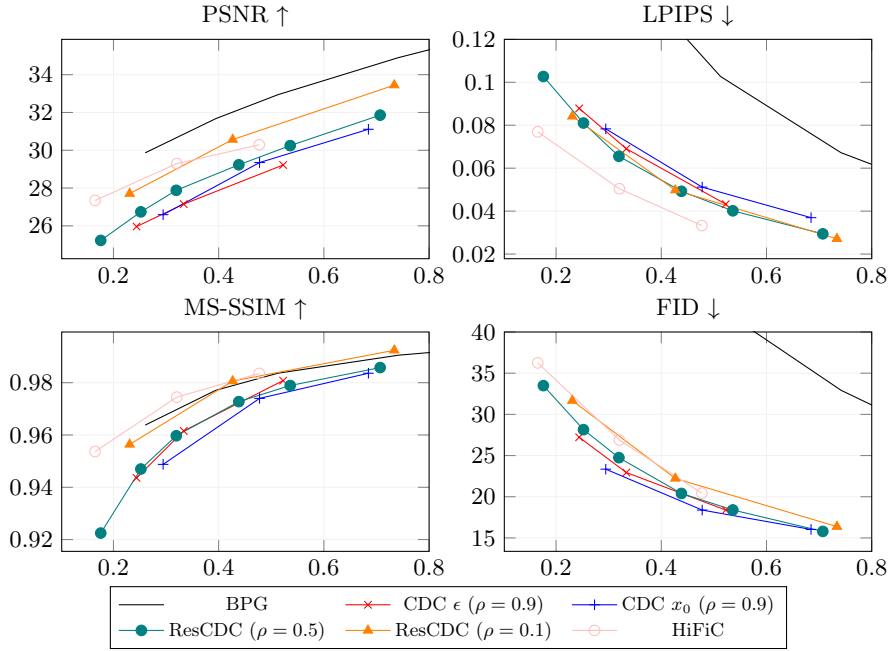
\subsection{Denoising Diffusion Probabilistic Models}
DDPMs~\cite{hoDenoisingDiffusionProbabilistic2020} work by gradually removing noise, until a fully denoised image is generated. 
For this purpose we model the forward diffusion process, which iteratively adds noise to an image according to some variance schedule, as a Markov chain.
By training a neural network to model the reverse process, iteratively removing noise from their input, diffusion models can produce high quality images.

While there are other approaches~\cite{delbracioInversionDirectIteration2023,bansalColdDiffusionInverting2022}, adding Gaussian noise has some advantages, as it allows us to directly generate a noisy image at an arbitrary timestep.
Given some noise schedule, with $\beta_t$ being the noise added at step $t$, the forward process $q(x_t | x_{t-1}) := \mathcal{N}(x_t; \sqrt{1-\beta_t}x_{t-1}, \beta_t \mathbf{I})$ can be simplified as follows \cite{hoDenoisingDiffusionProbabilistic2020}:
\begin{equation}
    q(x_t|x_0) = \mathcal{N}\big(x_t; \sqrt{\alpha_t}x_0, (1-\alpha_t) \mathbf{I}\big)\label{eq:forward}
\end{equation}
Where $\alpha_t := \prod^t_{s=1} (1 - \beta_s)$. In this work, we stick with a linear schedule, following the original DDPM paper~\cite{hoDenoisingDiffusionProbabilistic2020}. However, in some cases other approaches, such as a cosine based variance schedule, were found to be beneficial~\cite{nicholImprovedDenoisingDiffusion2021}. 
Note that if $\alpha_T \approx 0$ then $q(x_T | x_0) = \mathcal{N}(x_T; 0, \mathbf{I})$, which gives the natural starting point for the sampling process.

For the reverse process we use the neural network to obtain $p(x_{t-1} | x_t)$, which can be modelled in a few different ways. Either the model $f$ can be trained to directly predict $x_0$, giving the following formulation:
\begin{equation}
    p(x_{t-1} | x_t) = \mathcal{N}\big(x_{t-1}; \sqrt{\alpha_{t-1}} \cdot f(x_t, t, c), (1-\alpha_{t-1}) \mathbf{I}\big)
\end{equation}
Diffusion models can also be trained to predict the added noise $\epsilon$ instead of predicting $x_0$ directly, which was found to usually be easier to train \cite{hoDenoisingDiffusionProbabilistic2020}. 
Alternatively, $v$-prediction can be used which was introduced by Salimans and Ho \cite{salimans2021progressive} and is often found to lead to faster convergence. 
For $v$-prediction the network predicts $v = \sqrt{\alpha_t}  \epsilon - \sqrt{1 -\alpha_t}  x_0$.

In this paper we opted for this $\epsilon$ training objective for image compression, and $v$-prediction for the video experiments.

In either case the model is trained by sampling the timestep from $t \sim U(0, T)$, calculating the noisy image $x_t = \sqrt{\alpha_t}x_0 + \sqrt{1-\alpha_t}  \epsilon$ with $\epsilon \sim \mathcal{N}(0, \mathbf{I})$ and comparing the model output ($\hat x_0$, $\hat\epsilon$ or $\hat v$) to either the input image $x$, the added noise $\epsilon$, or $v$ respectively.

\subsection{Compression model training}
The diffusion model is trained to predict $\epsilon$ using the default simplified training objective $\mathcal{L}_{dist} =  \lvert\lvert\epsilon - \hat{\epsilon}\rvert\rvert_1$ as detailed in the DDPM paper \cite{hoDenoisingDiffusionProbabilistic2020}. Following existing works, we use the $l_1$ loss (instead of $l_2$) which was found to lead to fewer color artifacts \cite{yang2024lossy, saharia2022image}.

Additionally, we utilize a perceptual LPIPS~\cite{zhang2018unreasonable} loss component based on the VGG~\cite{simonyan2014very} network $\mathcal{L}_{perc} = LPIPS(x, \hat{x} + \hat{r}_0)$, similar to \cite{yang2024lossy, mentzerHighFidelityGenerativeImage2020}. Here $\hat{x}$ refers to the initial reconstruction of the decoder network and $\hat{r}_0$ refers to the one-step prediction of the residual by the diffusion model. The balance between the distortion and perceptual losses is controlled via the $\rho$ parameter.

While the decoder can be trained implicitly using the default diffusion training objective, similar to \cite{whang2022deblurring}, we can further steer the training by adding a $l_2$ loss component $\mathcal{L}_{decoder}=\lvert \lvert x - \hat{x}\rvert\rvert_2^2$ specifically for the decoder. We found this to be especially helpful when optimizing for perceptual quality.

Additionally, the bitrate loss $\mathcal{L}_{bitrate} = \log P(\hat{y}) + \log P(\hat{z})$ is required to train the encoder, similar to existing works for compression \cite{balleVariationalImageCompression2018, minnenJointAutoregressiveHierarchical2018, mentzerHighFidelityGenerativeImage2020}. 
Here $\hat{y}$ and $\hat{z}$ refer to the pseudo-quantized latents of the encoder. For inference the encoder latents $y$ (and hyper-latents $z$) are rounded to the next integer as follows: $\hat{y}_{inference} = \lfloor y\rceil$, which allows the use of arithmetic coding for compression.
However, as detailed in \cite{balleVariationalImageCompression2018}, to ensure we have proper gradients for learning, we use an approximation of this quantization during training. 
During training, we use the following definition $\hat{y} = \mathcal{U}(y - \frac{1}{2}, y + \frac{1}{2})$.

This gives us the following loss-function with two parameters $\lambda$ and $\rho$ to control the bitrate and the perceptual quality, respectively.
\begin{equation}
    \mathcal{L} =  \lambda \cdot \mathcal{L}_{bitrate} + (1 - \rho) \cdot \mathcal{L}_{dist} + \rho \cdot \mathcal{L}_{perc}\label{eq:loss} + \mathcal{L}_{decoder}
\end{equation}

For training, we sample random timesteps $t\sim \mathcal{U}(0, N)$ and generate the noisy samples $x_t$ according to Equation \eqref{eq:forward}.
The complete training procedure is shown in Algorithm \ref{alg:training}.

\subsection{Decompression / Sampling}

For sampling we use DDIM~\cite{songDenoisingDiffusionImplicit2022} which allows us to sample the model with significantly fewer steps.
In order to do so, Song \etal\cite{songDenoisingDiffusionImplicit2022} relax the Markov condition, in order to skip some of the steps during the reverse process. Another advantage of the DDIM sampling scheme, is that it is, apart from the starting point $x_T$, fully deterministic.
When setting $x_T=0$ the procedure even becomes fully deterministic.

We begin sampling at the starting point $x_T \sim \mathcal{N}(0, \mathbf{\gamma \cdot I})$. Similar as observed by Yang and Mandt for their CDC model~\cite{yang2024lossy}, we achieved the best results when setting $\gamma=0.8$ for models finetuned for perceptual quality. 
Using the reverse process, we iteratively denoise the image until $t=0$.

While the CDC noise-predicting diffusion model uses around 500 decoding steps, we evaluate our residual-based approach using only 100 decoding steps.

Note, that Yang \etal were able to achieve comparable results with only 17 decoding steps, when using a model which predicts $x_0$ instead of the noise $\epsilon$. 
While our model performs better on perceptual scores when using more steps, decoding with only 17 steps still gives good reconstructions. We further evaluate the effect of sampling steps in Section \ref{sec:sampling}

The complete decompression scheme is shown in Algorithm \ref{alg:sampling}.

\pgfplotscreateplotcyclelist{sampling}{
    thick, gray\\%
    orange\\%
    green\\%
    black\\%
    dashdotted, black\\%
    dashed, black!50\\%
    dotted, black!50\\%
    red,mark=x\\%
    blue,mark=x\\%
}
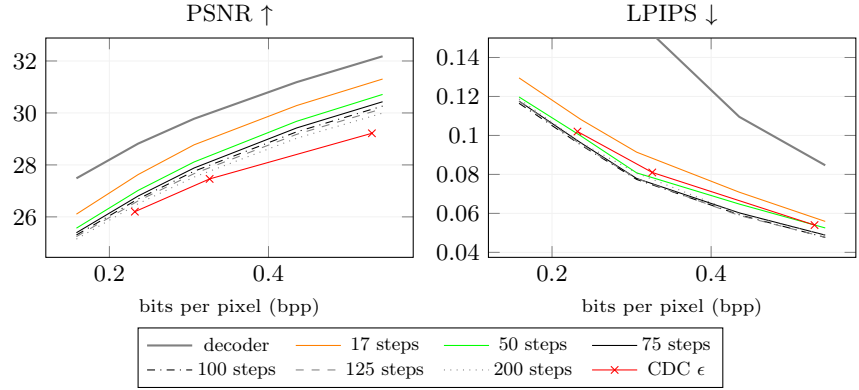
\begin{figure*}[!htb]
    \centering
    \footnotesize
    \begin{tikzpicture}
        \begin{groupplot}[
                legend style={at={(0.5,1.01)}, anchor=south, nodes={scale=.8, transform shape}},
                grid=both,
                grid style={line width=.1pt, draw=gray!10},
                group style={group size=2 by 2},
                legend columns=4,
                cycle list name=sampling,
                title style={yshift=-0.5em}, 
                xlabel={bits per pixel (bpp)},
                legend to name=leg-res-sample,
                width=.53\textwidth
            ]
            \nextgroupplot[
                title=PSNR \textuparrow, 
            ]
            
            \addplot table [x=bpp, y=psnr, col sep=comma] {csv/decode_steps/cdc_yang_orig_yang_orig_yang32_dmse1.0_residual_lpips_rho0.5_ft128_ema_decoder.csv};
            \addplot table [x=bpp, y=psnr, col sep=comma] {csv/decode_steps/cdc_yang_orig_yang_orig_yang32_dmse1.0_residual_lpips_rho0.5_ft128_ema_17.csv};
            \addplot table [x=bpp, y=psnr, col sep=comma] {csv/decode_steps/cdc_yang_orig_yang_orig_yang32_dmse1.0_residual_lpips_rho0.5_ft128_ema_50.csv};
            \addplot table [x=bpp, y=psnr, col sep=comma] {csv/decode_steps/cdc_yang_orig_yang_orig_yang32_dmse1.0_residual_lpips_rho0.5_ft128_ema_75.csv};
            \addplot table [x=bpp, y=psnr, col sep=comma] {csv/decode_steps/cdc_yang_orig_yang_orig_yang32_dmse1.0_residual_lpips_rho0.5_ft128_ema_100.csv};
            \addplot table [x=bpp, y=psnr, col sep=comma] {csv/decode_steps/cdc_yang_orig_yang_orig_yang32_dmse1.0_residual_lpips_rho0.5_ft128_ema_125.csv};
            \addplot table [x=bpp, y=psnr, col sep=comma] {csv/decode_steps/cdc_yang_orig_yang_orig_yang32_dmse1.0_residual_lpips_rho0.5_ft128_ema_200.csv};
            \addplot table [x=BPP, y=PSNR, col sep=comma] {csv/cdc_epsilon_rho09.csv};

            \nextgroupplot[
                title=LPIPS \textdownarrow, 
                ymax=0.15,
                ytick distance=0.02,
            ]            
            \addplot table [x=bpp, y=lpips, col sep=comma] {csv/decode_steps/cdc_yang_orig_yang_orig_yang32_dmse1.0_residual_lpips_rho0.5_ft128_ema_decoder.csv};
            \addplot table [x=bpp, y=lpips, col sep=comma] {csv/decode_steps/cdc_yang_orig_yang_orig_yang32_dmse1.0_residual_lpips_rho0.5_ft128_ema_17.csv};
            \addplot table [x=bpp, y=lpips, col sep=comma] {csv/decode_steps/cdc_yang_orig_yang_orig_yang32_dmse1.0_residual_lpips_rho0.5_ft128_ema_50.csv};
            \addplot table [x=bpp, y=lpips, col sep=comma] {csv/decode_steps/cdc_yang_orig_yang_orig_yang32_dmse1.0_residual_lpips_rho0.5_ft128_ema_75.csv};
            \addplot table [x=bpp, y=lpips, col sep=comma] {csv/decode_steps/cdc_yang_orig_yang_orig_yang32_dmse1.0_residual_lpips_rho0.5_ft128_ema_100.csv};
            \addplot table [x=bpp, y=lpips, col sep=comma] {csv/decode_steps/cdc_yang_orig_yang_orig_yang32_dmse1.0_residual_lpips_rho0.5_ft128_ema_125.csv};
            \addplot table [x=bpp, y=lpips, col sep=comma] {csv/decode_steps/cdc_yang_orig_yang_orig_yang32_dmse1.0_residual_lpips_rho0.5_ft128_ema_200.csv};
            \addplot table [x=BPP, y=LPIPS, col sep=comma] {csv/cdc_epsilon_rho09.csv};
            
            \legend{\footnotesize decoder, 17 steps, 50 steps, 75 steps, 100 steps, 125 steps, 200 steps, CDC $\epsilon$, CDC $\x_0$}
        \end{groupplot}
    \end{tikzpicture}
    \vspace{0cm}
    \pgfplotslegendfromname{leg-res-sample}
    \caption{Ablation comparing the impact of the amount of sampling steps on the Kodak images. The reference CDC $\epsilon$ was sampled using 500 steps. }\label{fig:eval:sampling}
\end{figure*}
\pgfplotscreateplotcyclelist{rho}{
    black\\%
    black!50\\%
    dashed\\%
    dotted\\%
    red,mark=x\\%
    blue,mark=+\\%
}
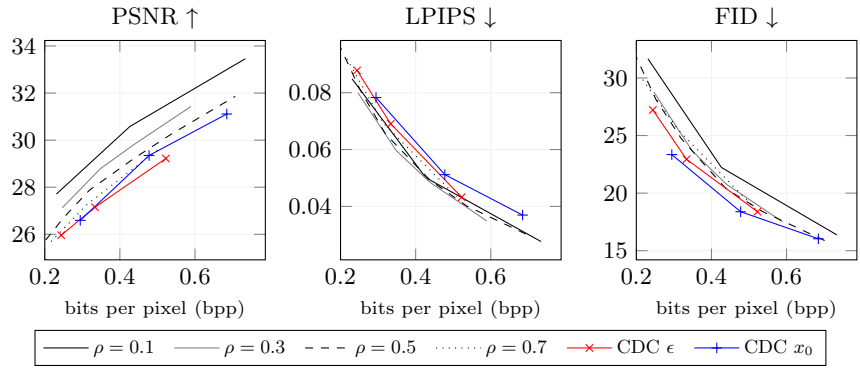
\begin{figure*}[!htb]
    \centering
    \footnotesize
    \begin{tikzpicture}
        \begin{groupplot}[
                legend style={at={(0.5,1.01)}, anchor=south, nodes={scale=.8, transform shape}},
                grid=both,
                grid style={line width=.1pt, draw=gray!10},
                group style={group size=3 by 2},
                legend columns=-1,
                cycle list name=rho,
                legend to name=leg-res-rho,
                width=.37\textwidth,
                xmin=0.2,
                title style={yshift=-0.5em}, 
                xlabel={bits per pixel (bpp)},
            ]
            \nextgroupplot[
                title=PSNR \textuparrow, 
            ]
            \addplot table [x=bpp, y=psnr, col sep=comma] {csv/div2k/rho_ablations/cdc_yang_orig_yang_orig_yang32_dmse1.0_residual_lpips_rho0.1_ft128_ema_100.csv};
            \addplot table [x=bpp, y=psnr, col sep=comma] {csv/div2k/rho_ablations/cdc_yang_orig_yang_orig_yang32_dmse1.0_residual_lpips_rho0.3_ft128_ema_100.csv};
            \addplot table [x=bpp, y=psnr, col sep=comma] {csv/div2k/rho_ablations/cdc_yang_orig_yang_orig_yang32_dmse1.0_residual_lpips_rho0.5_ft128_ema_100.csv};
            \addplot table [x=bpp, y=psnr, col sep=comma] {csv/div2k/rho_ablations/cdc_yang_orig_yang_orig_yang32_dmse1.0_residual_lpips_rho0.7_ft128_ema_100.csv};
            \addplot table [x=BPP, y=PSNR, col sep=comma] {csv/div2k/cdc_epsilon_rho09.csv};
            \addplot table [x=BPP, y=PSNR, col sep=comma] {csv/div2k/cdc_x0_rho09.csv};

            \nextgroupplot[
                title=LPIPS \textdownarrow, 
            ]     
            \addplot table [x=bpp, y=lpips, col sep=comma] {csv/div2k/rho_ablations/cdc_yang_orig_yang_orig_yang32_dmse1.0_residual_lpips_rho0.1_ft128_ema_100.csv};
            \addplot table [x=bpp, y=lpips, col sep=comma] {csv/div2k/rho_ablations/cdc_yang_orig_yang_orig_yang32_dmse1.0_residual_lpips_rho0.3_ft128_ema_100.csv};
            \addplot table [x=bpp, y=lpips, col sep=comma] {csv/div2k/rho_ablations/cdc_yang_orig_yang_orig_yang32_dmse1.0_residual_lpips_rho0.5_ft128_ema_100.csv};
            \addplot table [x=bpp, y=lpips, col sep=comma] {csv/div2k/rho_ablations/cdc_yang_orig_yang_orig_yang32_dmse1.0_residual_lpips_rho0.7_ft128_ema_100.csv};
            \addplot table [x=BPP, y=LPIPS, col sep=comma] {csv/div2k/cdc_epsilon_rho09.csv};
            \addplot table [x=BPP, y=LPIPS, col sep=comma] {csv/div2k/cdc_x0_rho09.csv};
            
            \nextgroupplot[
                title=FID \textdownarrow, 
            ]     
            \addplot table [x=bpp, y=fid, col sep=comma] {csv/div2k/rho_ablations/cdc_yang_orig_yang_orig_yang32_dmse1.0_residual_lpips_rho0.1_ft128_ema_100.csv};
            \addplot table [x=bpp, y=fid, col sep=comma] {csv/div2k/rho_ablations/cdc_yang_orig_yang_orig_yang32_dmse1.0_residual_lpips_rho0.3_ft128_ema_100.csv};
            \addplot table [x=bpp, y=fid, col sep=comma] {csv/div2k/rho_ablations/cdc_yang_orig_yang_orig_yang32_dmse1.0_residual_lpips_rho0.5_ft128_ema_100.csv};
            \addplot table [x=bpp, y=fid, col sep=comma] {csv/div2k/rho_ablations/cdc_yang_orig_yang_orig_yang32_dmse1.0_residual_lpips_rho0.7_ft128_ema_100.csv};
            \addplot table [x=BPP, y=FID, col sep=comma] {csv/div2k/cdc_epsilon_rho09.csv};
            \addplot table [x=BPP, y=FID, col sep=comma] {csv/div2k/cdc_x0_rho09.csv};
            
            \legend {$\rho=0.1$, $\rho=0.3$, $\rho=0.5$, $\rho=0.7$, CDC $\epsilon$, CDC $x_0$}
        \end{groupplot}
    \end{tikzpicture}
    \vspace{0cm}
    \pgfplotslegendfromname{leg-res-rho}
    \caption{Ablation for the perceptual-trade-off parameter $\rho$ on the DIV2K dataset. ResCDC sampled with 100 steps. CDC baselines sampled with 500 and 17 steps respectively.} 
    \label{fig:eval:rho}
\end{figure*}
\section{Experiments}

In the following we describe our experimental setup, give implementation details and provide results and ablations on various datasets and standard metrics.

\subsection{Experimental setup}
\noindent\textbf{Datasets}
Following \cite{yang2024lossy} we provide results for these datasets:
\textbf{Kodak}~\cite{kodak}, which contains 24 images with resolutions of 512$\times$768 or 768$\times$512. 
As well as the \textbf{DIV2K}~\cite{Agustsson_2017_CVPR_Workshops} validation-split which contains 100 high-resolution images, which we resize and crop to a resolution of 768$\times$768 (following \cite{yang2024lossy}).

\noindent\textbf{Measures}
We evaluate our approach using PSNR, MS-SSIM~\cite{wang2003multiscale}, LPIPS~\cite{zhang2018unreasonable} (using the AlexNet backend) and FID~\cite{heusel2017gans}, which are popular metrics for evaluating distortion and perceptual quality. 
For evaluating the FID we split the images into 256$\times$256 image patches similar to previous works
\cite{mentzerHighFidelityGenerativeImage2020}. Since the Kodak dataset only consists of 24 images, we do not have enough samples to calculate representative FID scores. Therefore, we only report FID for the DIV2K dataset. 

\noindent\textbf{Compared methods}
We compare our results against the original perceptual CDC models~\cite{yang2024lossy}, as well as the standard BPG~\cite{bpg}, which was included as a fidelity-optimized reference and HiFiC~\cite{mentzerHighFidelityGenerativeImage2020}, as a perceptual-optimized reference. Similar to fidelity-optimized learned image methods, BPG does not score well on perceptual metrics. \footnote{HiFiC Pytorch reimplmentation: \url{https://github.com/Justin-Tan/high-fidelity-generative-compression}}
While we only provide a model trained to predict the noise $\epsilon$, the original CDC also provides results for a diffusion model trained to predict the image $x_0$.

\subsection{Implementation details}
Following the work of Yang \etal\cite{yang2024lossy} our models are trained using the Vimeo90k-septuplet~\cite{xue2019video} dataset, which consists of roughly 90,000 7-frame sequences with a fixed resolution of 448$\times$256. We select a random frame of every sequence and crop the image randomly to a resolution of 256$\times$256.

For model training we use the Adam optimizer with a learning rate of $5e-5$, which declines by $1e-5$ every 100k steps until the learning rate reaches $2e-5$. The model is trained on a single GPU, with batchsize 4.
The rate parameter is initialized at $\lambda = 1e-5$ for a 500k step warm-up, after which $\lambda$ is set according to the desired bitrate.
The models are then pretrained with $\rho=0$ for 1M steps, after which we finetune for perceptual quality for another 1M steps using $\rho>0$.

Our implementation is based on the original implementation of the CDC~\cite{yang2024lossy} model. 
We expect overall training and testing complexity to be comparable. 
Our implementation is available on Github\footnote{\url{https://github.com/jbrenig/ResCDC}}.

\subsection{Comparison with state-of-the-art}
As shown in Figure~\ref{fig:eval:kodak}, for the Kodak images our approach (ResCDC) generally produces images with similar or slightly better perceptual quality, while at the same time scoring higher on PSNR (up to +2dB).
The same holds for the DIV2K dataset (as shown in Figure~\ref{fig:eval:div2k}). Here we also report FID in which our approach performs comparable or, in some cases, slightly worse than CDC~\cite{yang2024lossy}. Similar to the original CDC, our method gets outperformed by HiFiC~\cite{mentzerHighFidelityGenerativeImage2020} in terms of LPIPS, while still improving over HiFiC in metrics such as FID.

Comparing with CDC-$\epsilon$ in terms of BD-Rate~\cite{bjontegaard2001calculation} for PSNR we see an improvement of $-13.6\%$ on Kodak and $-17.4\%$ on Div2k (for ResCDC $\rho=0.5$).

We show results for two variants, one tuned more for perceptual quality ($\rho=0.5$), the other tuned for better distortion ($\rho=0.1$). While both variants score similar in LPIPS, there is a trade-off between FID and distortion metrics. Note that both variants score higher in PSNR compared to CDC. We provide ablations for different $\rho$ in Section \ref{sec:rho}.

While the decoder on its own scores best in PSNR (see Figure \ref{fig:eval:sampling}), it is still significantly worse than even hand-crafted codecs such as BPG. 
However, even a low amount of sampling steps of the diffusion model drastically improves perceptual quality.

As shown in the qualitative comparison in Figure \ref{fig:qualitative}, the initial decoder reconstruction gets refined by the diffusion model, adding additional detail.
Compared to CDC, our model generally generates images that are closer to the original image, and are often sharper. In some cases this comes at the cost of smaller artifacts or image noise.

\subsection{Ablative studies}
\subsubsection{Number of sampling steps}\label{sec:sampling}
As detailed in Figure~\ref{fig:eval:sampling}, our approach performs best around 100 sampling steps. 
When decoding fewer sampling steps we see improvements to PSNR at the cost of perceptual quality. This is inline with the findings of Blau and Michaeli that there always exists a trade-off between perceptual quality and distortion measures \cite{blau2018perception}. 

While the original CDC paper only requires 17 steps for decoding using the $x_0$-predicting model, our ($\epsilon$-predicting) model achieves only slightly worse LPIPS scores, while showing much higher PSNR when using the same number of decoding steps.

\subsubsection{Effect of the $\rho$ parameter}\label{sec:rho}
Since in our case the diffusion model works on the residual, the optimal values for $\rho$ are much lower compared to the base CDC model.
As detailed in Figure \ref{fig:eval:rho}, higher values for $\rho$ generally improve perceptual quality at the cost of PSNR. While lower values for $\rho$ can still score similar on LPIPS, other perceptual metrics such as FID suffer.
For our main results we find $\rho=0.5$ to be a good compromise. At this setting our model shows a significant improvement for PSNR, with only small impact on FID scores. We also report results for $\rho=0.1$, a setting with significantly better PSNR scores at the cost of lower FID scores.

For values $\rho > 0.7$ the generated image generally decreased in all evaluated metrics. Similarly, values $\rho < 0.1$ lead to much lower perceptual quality, while increases in PSNR taper off eventually.

At higher bitrates the advantage of ResCDC is even more apparent, as even models with lower $\rho$ values show good scores in perceptual metrics and an even larger lead in PSNR.

\subsection{ResCDC for Video Compression}\label{sec:video}
Our approach can also easily be adapted for learned video compression methods and therefore mostly follows the same architecture and training procedure as the image compression model.
In order to handle video, we modify the diffusion unet to include the current decoder reconstruction, as well as the previously reconstructed frame as additional conditioning. During training we use the decoder reconstruction as the previous frame conditioning, while during testing we can use the higher quality diffusion result of the previous frame.

The encoder-decoder backbone is replaced by a pretrained learned video compression method. We train the diffusion model on top of a pretrained backbone (instead of training everything from scratch).
The sampling speed of diffusion models is critical when applied for videos (especially in the context of compression). Therefore, we evaluate the diffusion model using only 10 sampling steps. While a higher amount of sampling steps can improve the final results, it also significantly increases the time needed to decode the video. However, decoding remains slow and we are mostly interested in potential perceptual quality improvements.
We found $\gamma = 0.1$ to give the best results when only using 10 sampling steps, with higher values for $\gamma$ producing noisy outputs.

Since most Video Compression methods use different models for the keyframes and progressive frames, we also train separate diffusion models for both. 

In contrast to the image compression model, we opted to use $v$-prediction for the video compression diffusion model.

\noindent\textbf{Training} The diffusion model is trained on top of the pretrained video compression method using the Vimeo90k septuplet dataset~\cite{xue2019video}. We randomly select three consecutive frames of the septuplet and train the diffusion model on top of a frozen video compression method, before finetuning the whole model.

\noindent\textbf{Compared methods} Our method can theoretically be applied on top of any learned video compression method with available latent information. Due to high training cost we only provide results of our ResCDC model trained on top of the Scale-Space Flow~\cite{agustsson2020scale} and DVC~\cite{lu2019dvc} compression methods.  
We compare our results against the baseline Scale-Space Flow~\cite{agustsson2020scale} and DVC~\cite{lu2019dvc} models, as well as the more recent DCVC-FM~\cite{li2021deep, li2024neural}.

We use pretrained checkpoints provided by open-source PyTorch implementations of said methods. In the case of DVC\footnote{DVC: \url{https://github.com/ZhihaoHu/PyTorchVideoCompression}} and Scale-Space Flow\footnote{SSF2020: \url{https://github.com/InterDigitalInc/CompressAI}} we resort to re-implementations (which might not reach the same performance as the originals), while for DVCV-FM\footnote{DCVC-FM: \url{https://github.com/microsoft/DCVC}} we use the implementation by the original authors.

\noindent\textbf{Datasets} We evaluate our video compression model on the UVG dataset~\cite{mercat2020uvg}, consisting of seven high-resolution video clips of 300 or 600 frames. Following existing literature we set the intra-frame period to 9 frames. Note that DCVC-FM~\cite{li2024neural} is optimized to run on much higher intra-frame periods. However, other methods see severely degraded performance in this case.

\pgfplotscreateplotcyclelist{video}{
    red,mark=x\\%
    red,mark=otimes\\%
    orange,mark=*\\%
    black\\%
    black,mark=square\\%
    blue,mark=square*\\%
    teal,mark=*\\%
    orange,mark=*\\%
    pink,mark=*\\%
    pink,mark=*\\%
    pink,mark=*\\%
}
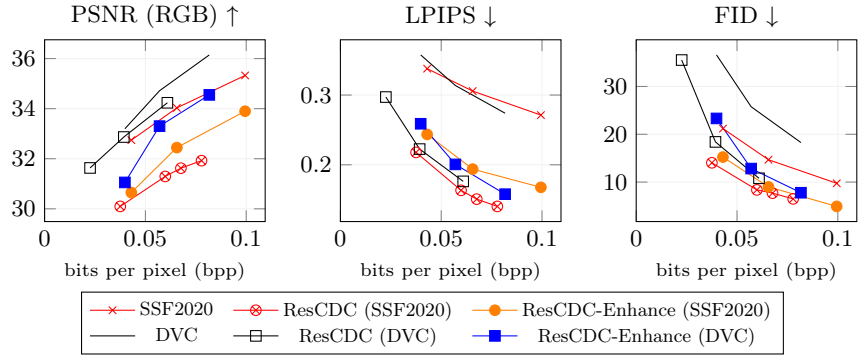
\begin{figure*}[!htb]
    \centering
    \footnotesize
    \begin{tikzpicture}
        \begin{groupplot}[
                legend style={at={(0.5,1.01)}, anchor=south, nodes={scale=0.8, transform shape}},
                grid=both,
                grid style={line width=.1pt, draw=gray!10},
                group style={group size=3 by 2},
                legend columns=3,
                cycle list name=video,
                xlabel={bits per pixel (bpp)},
                legend to name=leg-res-vid,
                width=.37\textwidth,
                title style={yshift=-0.5em}, 
                xmin=0.0,
                height=4cm,
            ]
            \nextgroupplot[
                title=PSNR (RGB) \textuparrow, 
            ]%
            \addplot table [x=bpp, y=psnr, col sep=comma] {csv/video/pretrained-ssf2020.csv};

            \addplot table [x=bpp, y=psnr, col sep=comma] {csv/video/vrescdc_vimeo90k-septuplet_ft-q__ssf2020_unet=yang_orig_lfc_cfc_yang_orig_v.csv};
            \addplot table [x=bpp, y=psnr, col sep=comma] {csv/video/vrescdc_vimeo90k-septuplet_enhance_rho1__ssf2020_unet=yang_orig_freeze_ae_lfc_cfc_yang_orig_v.csv};

            \addplot table [x=bpp, y=psnr, col sep=comma] {csv/video/pretrained-dvc.csv};
            \addplot table [x=bpp, y=psnr, col sep=comma] {csv/video/vrescdc_vimeo90k-septuplet_v_ft__dvc_mbt2018-mean_unet=yang_orig_lfc_cfc_dual_unet_yang_orig_v.csv};
            \addplot table [x=bpp, y=psnr, col sep=comma] {csv/video/vrescdc_vimeo90k-septuplet_v_ft__dvc_mbt2018-mean_unet=yang_orig_freeze_ae_lfc_cfc_dual_unet_yang_orig_v.csv};

            \nextgroupplot[
                title=LPIPS \textdownarrow, 
            ]%
            \addplot table [x=bpp, y=lpips, col sep=comma] {csv/video/pretrained-ssf2020.csv};
            
            \addplot table [x=bpp, y=lpips, col sep=comma] {csv/video/vrescdc_vimeo90k-septuplet_ft-q__ssf2020_unet=yang_orig_lfc_cfc_yang_orig_v.csv};
            \addplot table [x=bpp, y=lpips, col sep=comma] {csv/video/vrescdc_vimeo90k-septuplet_enhance_rho1__ssf2020_unet=yang_orig_freeze_ae_lfc_cfc_yang_orig_v.csv};

            \addplot table [x=bpp, y=lpips, col sep=comma] {csv/video/pretrained-dvc.csv};
            \addplot table [x=bpp, y=lpips, col sep=comma] {csv/video/vrescdc_vimeo90k-septuplet_v_ft__dvc_mbt2018-mean_unet=yang_orig_lfc_cfc_dual_unet_yang_orig_v.csv};
            \addplot table [x=bpp, y=lpips, col sep=comma] {csv/video/vrescdc_vimeo90k-septuplet_v_ft__dvc_mbt2018-mean_unet=yang_orig_freeze_ae_lfc_cfc_dual_unet_yang_orig_v.csv};
        
            \nextgroupplot[
                title=FID \textdownarrow, 
            ]%
            \addplot table [x=bpp, y=fid, col sep=comma] {csv/video/pretrained-ssf2020.csv};
            
            \addplot table [x=bpp, y=fid, col sep=comma] {csv/video/vrescdc_vimeo90k-septuplet_ft-q__ssf2020_unet=yang_orig_lfc_cfc_yang_orig_v.csv};
            \addplot table [x=bpp, y=fid, col sep=comma] {csv/video/vrescdc_vimeo90k-septuplet_enhance_rho1__ssf2020_unet=yang_orig_freeze_ae_lfc_cfc_yang_orig_v.csv};

            \addplot table [x=bpp, y=fid, col sep=comma] {csv/video/pretrained-dvc.csv};
            \addplot table [x=bpp, y=fid, col sep=comma] {csv/video/vrescdc_vimeo90k-septuplet_v_ft__dvc_mbt2018-mean_unet=yang_orig_lfc_cfc_dual_unet_yang_orig_v.csv};
            \addplot table [x=bpp, y=fid, col sep=comma] {csv/video/vrescdc_vimeo90k-septuplet_v_ft__dvc_mbt2018-mean_unet=yang_orig_freeze_ae_lfc_cfc_dual_unet_yang_orig_v.csv};
            
            \legend {SSF2020, ResCDC (SSF2020), ResCDC-Enhance (SSF2020), DVC, ResCDC (DVC), ResCDC-Enhance (DVC)}
        \end{groupplot}
    \end{tikzpicture}
    \vspace{0cm}
    \pgfplotslegendfromname{leg-res-vid}
    \caption{Results for using ResCDC for video \emph{compression} on the UVG dataset. ResCDC models are sampled with 10 steps. ResCDC-Enhance refers to a post-processing model trained on a frozen backend.}
    \label{fig:eval:vid}
\end{figure*}

\noindent\textbf{Results} As seen in Figure~\ref{fig:eval:vid}, our method significantly improves perceptual quality compared to the baseline codecs. However, similar to the image compression results, we see a significant drop in PSNR, dropping more than 2dB in some settings.
In most cases, we did not notice problems with temporal consistency compared to the baseline codecs. However, for low bitrates some inconsistent colors between the keyframes and progressive were noticeable. This might be caused by the use of two distinct networks, instead of a single diffusion model.

\noindent\textbf{Latent conditioned compressed video enhancement} 
It is also possible to skip the end-to-end training, and to use the diffusion model only as an enhancement step of an existing neural codec. As seen in Figure~\ref{fig:eval:vid}, this approach leads to slightly worse perceptual quality compared to the end-to-end trained model. However, in the case of Scale-Space Flow, the enhancement model actually performed slightly better in regards to PSNR.

\begin{figure*}[!htb]
    \centering
    \footnotesize
    \begin{tikzpicture}
        \begin{groupplot}[
                legend style={at={(0.5,1.01)}, anchor=south, nodes={scale=0.8, transform shape}},
                grid=both,
                grid style={line width=.1pt, draw=gray!10},
                group style={group size=3 by 2},
                legend columns=3,
                cycle list name=videnhance,
                xlabel={bits per pixel (bpp)},
                legend to name=leg-res-vid-enh,
                width=.37\textwidth,
                title style={yshift=-0.5em}, 
                xmin=0.0,
                height=4cm,
            ]
            \nextgroupplot[
                title=PSNR (RGB) \textuparrow
            ]
            \addplot table [x=bpp, y=psnr, col sep=comma] {csv/video/pretrained-ssf2020.csv};
            \addplot table [x=bpp, y=psnr, col sep=comma] {csv/video/pretrained-dcvc_fm.csv};
            \addplot table [x=bpp, y=psnr, col sep=comma] {csv/video/pretrained-dvc.csv};
            
            \addplot table [x=bpp, y=psnr, col sep=comma] {csv/video/vrescdc_vimeo90k-septuplet_enhance_rho1__ssf2020_unet=yang_orig_freeze_ae_lfc_cfc_yang_orig_v.csv};
            \addplot table [x=bpp, y=psnr, col sep=comma] {csv/video/vrescdc_vimeo90k-septuplet_enhance_rho1_ft__dcvc_fm_ifq63_pfq63_unet=yang_orig_freeze_ae_lfc_cfc_yang_orig_v.csv};
            \addplot table [x=bpp, y=psnr, col sep=comma] {csv/video/vrescdc_vimeo90k-septuplet_v_ft__dvc_mbt2018-mean_unet=yang_orig_freeze_ae_lfc_cfc_dual_unet_yang_orig_v.csv};
            \addplot table [x=bpp, y=psnr, col sep=comma] {csv/video/vrescdc_vimeo90k-septuplet_enhance_rho1_nlc__ssf2020_unet=yang_orig_freeze_ae_lfc_cfc_nlc_yang_orig_v.csv};
            \addplot table [x=bpp, y=psnr, col sep=comma] {csv/video/vrescdc_vimeo90k-septuplet_enhance_rho1_ft_nlc__dcvc_fm_ifq63_pfq63_unet=yang_orig_freeze_ae_lfc_cfc_yang_orig_v.csv};

            \nextgroupplot[title=LPIPS \textdownarrow]     
            \addplot table [x=bpp, y=lpips, col sep=comma] {csv/video/pretrained-ssf2020.csv};
            \addplot table [x=bpp, y=lpips, col sep=comma] {csv/video/pretrained-dcvc_fm.csv};
            \addplot table [x=bpp, y=lpips, col sep=comma] {csv/video/pretrained-dvc.csv};
            
            \addplot table [x=bpp, y=lpips, col sep=comma] {csv/video/vrescdc_vimeo90k-septuplet_enhance_rho1__ssf2020_unet=yang_orig_freeze_ae_lfc_cfc_yang_orig_v.csv};
            \addplot table [x=bpp, y=lpips, col sep=comma] {csv/video/vrescdc_vimeo90k-septuplet_enhance_rho1_ft__dcvc_fm_ifq63_pfq63_unet=yang_orig_freeze_ae_lfc_cfc_yang_orig_v.csv};
            \addplot table [x=bpp, y=lpips, col sep=comma] {csv/video/vrescdc_vimeo90k-septuplet_v_ft__dvc_mbt2018-mean_unet=yang_orig_freeze_ae_lfc_cfc_dual_unet_yang_orig_v.csv};
            \addplot table [x=bpp, y=lpips, col sep=comma] {csv/video/vrescdc_vimeo90k-septuplet_enhance_rho1_nlc__ssf2020_unet=yang_orig_freeze_ae_lfc_cfc_nlc_yang_orig_v.csv};
            \addplot table [x=bpp, y=lpips, col sep=comma] {csv/video/vrescdc_vimeo90k-septuplet_enhance_rho1_ft_nlc__dcvc_fm_ifq63_pfq63_unet=yang_orig_freeze_ae_lfc_cfc_yang_orig_v.csv};
            
            \nextgroupplot[title=FID \textdownarrow]     
            \addplot table [x=bpp, y=fid, col sep=comma] {csv/video/pretrained-ssf2020.csv};
            \addplot table [x=bpp, y=fid, col sep=comma] {csv/video/pretrained-dcvc_fm.csv};
            \addplot table [x=bpp, y=fid, col sep=comma] {csv/video/pretrained-dvc.csv};
            
            \addplot table [x=bpp, y=fid, col sep=comma] {csv/video/vrescdc_vimeo90k-septuplet_enhance_rho1__ssf2020_unet=yang_orig_freeze_ae_lfc_cfc_yang_orig_v.csv};
            \addplot table [x=bpp, y=fid, col sep=comma] {csv/video/vrescdc_vimeo90k-septuplet_enhance_rho1_ft__dcvc_fm_ifq63_pfq63_unet=yang_orig_freeze_ae_lfc_cfc_yang_orig_v.csv};
            \addplot table [x=bpp, y=fid, col sep=comma] {csv/video/vrescdc_vimeo90k-septuplet_v_ft__dvc_mbt2018-mean_unet=yang_orig_freeze_ae_lfc_cfc_dual_unet_yang_orig_v.csv};
            \addplot table [x=bpp, y=fid, col sep=comma] {csv/video/vrescdc_vimeo90k-septuplet_enhance_rho1_nlc__ssf2020_unet=yang_orig_freeze_ae_lfc_cfc_nlc_yang_orig_v.csv};
            \addplot table [x=bpp, y=fid, col sep=comma] {csv/video/vrescdc_vimeo90k-septuplet_enhance_rho1_ft_nlc__dcvc_fm_ifq63_pfq63_unet=yang_orig_freeze_ae_lfc_cfc_yang_orig_v.csv};

            \legend {SSF2020, DCVC-FM, DVC, ResCDC (SSF2020), ResCDC (DCVC-FM), ResCDC (DVC), ResCDC-NL (SSF2020), ResCDC-NL (DCVC-FM)}
        \end{groupplot}
    \end{tikzpicture}
    \vspace{0cm}
    \pgfplotslegendfromname{leg-res-vid-enh}
    \caption{Results for using ResCDC for compressed video \emph{enhancement} of neural codecs on the UVG dataset. ResCDC models are sampled with 10 steps. -NL denotes a model that does not use the encoder latent for conditioning.}
    \label{fig:eval:vid:enhance}
\end{figure*}
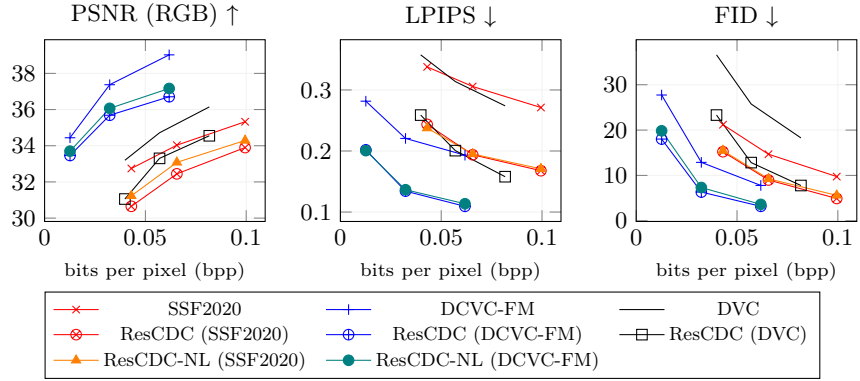

In Figure \ref{fig:eval:vid:enhance} we compare different configurations of using our method as an enhancement step to a pretrained neural compression codec. For these experiments we also show results for a ResCDC model trained on top of the DCVC-FM~\cite{li2024neural} model, which can be adjusted to different bitrates during testing time. 

In this setting, it is also possible to omit the encoder latent conditioning, meaning the diffusion model will only be conditioned on the decoder reconstructions. Results without the encoder latent for conditioning are indicated by ``-NL''.
Our experiments indicate that the latent does not make a relevant difference in image quality, when using ResCDC for video enhancement.
Therefore, in the case of video compression, most of the benefit of perceptual quality can be achieved by using a diffusion model as a post processing step.

\pgfplotscreateplotcyclelist{videnhance}{
    red,mark=x\\%
    blue,mark=+\\%
    black\\%
    red,mark=otimes\\%
    blue,mark=oplus\\%
    black,mark=square\\%
    orange,mark=triangle*\\%
    teal,mark=*\\%
}

\section{Conclusion}

In this paper we propose an end-to-end trained neural image compression model optimized for perceptual quality. 
Using a latent conditioned diffusion model to predict the residuals of the autoencoder reconstruction, this method significantly improves the fidelity of the reconstructed images compared to a diffusion-only approach. While the decoder networks generates an initial distortion-optimized reconstruction, the diffusion model adds additional details to generate images of higher perceptual quality. We were able to demonstrate significant gains in PSNR scores, up to +2dB,  at similar perceptual quality, especially at higher bitrates, when compared to CDC~\cite{yang2024lossy}.

By using an off-the-shelf decoder trained in combination with a diffusion model, a high-quality reconstruction can be obtained using only a few sampling steps. Still, decoding time is inherently slower compared to existing autoencoder based approaches, due to the iterative sampling procedure.

We show that our approach can also be applied to the domain of perceptual video compression. While the slow decoding speed is even more noticeable in this setting, we are able to finetune existing learned video compression methods for significantly improved perceptual quality at the cost of scoring lower on distortion metrics such as PSNR. 

\noindent\paragraph{Acknowledgments. }
This work was partly supported by the Alexander von Humboldt Foundation.

\bibliographystyle{splncs04}
\bibliography{main}

\clearpage






\appendix
\section{Additional experimental results}
In this section we compare the number of parameters for each model, and provide some additional results for other perceptual metrics. 
Generally our approach performs very similar to the base CDC model in most perceptual metrics.

\begin{table}[!htb]
    \centering
    \begin{tabular}{lr}
    \toprule
       Model  & \#Parameters \\
       \midrule
       MS-Hyper  & 17.5 M \\ 
       HiFiC  & 181.5 M \\ 
       CDC  & 53.8 M \\ 
       ResCDC  & 58.7 M \\
       \bottomrule\\
    \end{tabular}
    \caption{Number of parameters, Image models}
    \label{tab:params:image}
\end{table}
\begin{table}[!htb]
    \centering
    \begin{tabular}{lrr}
    \toprule
       Base-Model  & \#Param (Base) & \#Param (ResCDC) \\
       \midrule
       SSF ($q=3$)  & 34.2 M & 118.8 M \\ 
       DVC ($q=2$)  & 11.8 M & 96.1 M \\
       DCVC-FM      & 44.9 M & 129.3 M \\
       \bottomrule\\
    \end{tabular}
    \caption{Number of parameters, Video models}
    \label{tab:params:video}
\end{table}

\pgfplotscreateplotcyclelist{main}{
    red,mark=x\\%
    blue,mark=+\\%
    teal,mark=*\\%
    orange,mark=triangle*\\%
    pink,mark=o\\%
}

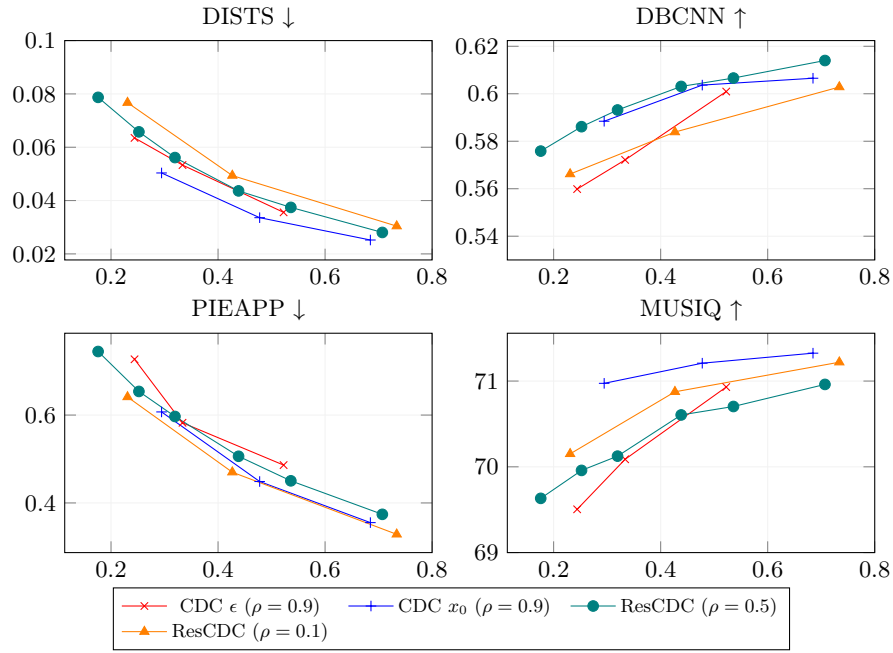
\begin{figure*}[!htb]
    \centering
    \begin{tikzpicture}
         \begin{groupplot}[
                legend style={at={(0.5,1.01)}, anchor=south, nodes={scale=0.8, transform shape}},
                grid=both,
                grid style={line width=.1pt, draw=gray!10},
                group style={
                    group size=2 by 2,
                    vertical sep=3em
                },
                legend columns=3,
                cycle list name=main,
                legend to name=leg-res-div2k,
                title style={yshift=-0.5em}, 
                width=.53\textwidth,
                xmax=0.8,
            ]
            \nextgroupplot[
                title=DISTS \textdownarrow,
                ymax=0.1
            ]           
            \addplot table [x=BPP, y=dists, col sep=comma, blue, mark=o] {csv/div2k/cdc_epsilon_rho09.csv};
            \addplot table [x=BPP, y=dists, col sep=comma] {csv/div2k/cdc_x0_rho09.csv};
            \addplot table [x=bpp, y=dists, col sep=comma] {csv/div2k/cdc_residual_epsilon_rho05.csv};
            \addplot table [x=bpp, y=dists, col sep=comma] {csv/div2k/cdc_residual_epsilon_rho01.csv};
            
            \nextgroupplot[
                title=DBCNN \textuparrow,
                ymin=0.53
            ]            
            \addplot table [x=BPP, y=dbcnn, col sep=comma] {csv/div2k/cdc_epsilon_rho09.csv};
            \addplot table [x=BPP, y=dbcnn, col sep=comma] {csv/div2k/cdc_x0_rho09.csv};
            \addplot table [x=bpp, y=dbcnn, col sep=comma] {csv/div2k/cdc_residual_epsilon_rho05.csv};
            \addplot table [x=bpp, y=dbcnn, col sep=comma] {csv/div2k/cdc_residual_epsilon_rho01.csv};
            
            \nextgroupplot[
                title=PIEAPP \textdownarrow,
            ]             
            \addplot table [x=BPP, y=piq-pieapp, col sep=comma] {csv/div2k/cdc_epsilon_rho09.csv};
            \addplot table [x=BPP, y=piq-pieapp, col sep=comma] {csv/div2k/cdc_x0_rho09.csv};
            \addplot table [x=bpp, y=piq-pieapp, col sep=comma] {csv/div2k/cdc_residual_epsilon_rho05.csv};
            \addplot table [x=bpp, y=piq-pieapp, col sep=comma] {csv/div2k/cdc_residual_epsilon_rho01.csv};

            \nextgroupplot[
                title=MUSIQ \textuparrow,
                ymin=69,
            ]            
            \addplot table [x=BPP, y=pyiqa-musiq, col sep=comma] {csv/div2k/cdc_epsilon_rho09.csv};
            \addplot table [x=BPP, y=pyiqa-musiq, col sep=comma] {csv/div2k/cdc_x0_rho09.csv};
            \addplot table [x=bpp, y=pyiqa-musiq, col sep=comma] {csv/div2k/cdc_residual_epsilon_rho05.csv};
            \addplot table [x=bpp, y=pyiqa-musiq, col sep=comma] {csv/div2k/cdc_residual_epsilon_rho01.csv};
            
            \legend {CDC $\epsilon$ ($\rho=0.9$), CDC $x_0$ ($\rho=0.9$), ResCDC ($\rho=0.5$),  ResCDC ($\rho=0.1$)}
        \end{groupplot}
    \end{tikzpicture}
    \vspace{0cm}
    \pgfplotslegendfromname{leg-res-div2k}
    \caption{Additional results on the DIV2K validation-set for various perceptual metrics. CDC $\epsilon$ models were sampled using 500 steps, CDC $x_0$ models using 17 steps, ResCDC (ours) models using 100 steps.}\label{fig:eval:div2k}
\end{figure*}


\end{document}